\documentclass[twocolumn]{aastex63}

\submitjournal{ApJ}

\shorttitle{SILICON AND HYDROGEN CHEMISTRY: MIMICKING AGBs CONDITIONS}
\shortauthors{Accolla et al.}

\begin{document}

\title{SILICON AND HYDROGEN CHEMISTRY UNDER LABORATORY CONDITIONS MIMICKING THE ATMOSPHERE OF EVOLVED STARS}

\correspondingauthor{Mario Accolla, Gonzalo Santoro, Jos\'e \'Angel Mart\'i{}n-Gago }
\email{mario.accolla@csic.es; gonzalo.santoro@icmm.csic.es; gago@icmm.csic.es}

\author{Mario Accolla}
\affiliation{Instituto de Ciencia de Materiales de Madrid (ICMM. CSIC). Materials Science Factory. Structure of Nanoscopic Systems Group. c/ Sor Juana In\'es de la Cruz 3, 28049 Cantoblanco, Madrid, Spain.}

\author{Gonzalo Santoro}
\affiliation{Instituto de Ciencia de Materiales de Madrid (ICMM. CSIC). Materials Science Factory. Structure of Nanoscopic Systems Group. c/ Sor Juana In\'es de la Cruz 3, 28049 Cantoblanco, Madrid, Spain.}

\author{Pablo Merino}
\affiliation{Instituto de Ciencia de Materiales de Madrid (ICMM. CSIC). Materials Science Factory. Structure of Nanoscopic Systems Group. c/ Sor Juana In\'es de la Cruz 3, 28049 Cantoblanco, Madrid, Spain.}
\affiliation{Instituto de F\'\i{}sica Fundamental (IFF. CSIC). Group of Molecular Astrophysics. c/ Serrano 123, 28006 Madrid, Spain.}

\author{Lidia Mart\'\i{}nez}
\affiliation{Instituto de Ciencia de Materiales de Madrid (ICMM. CSIC). Materials Science Factory. Structure of Nanoscopic Systems Group. c/ Sor Juana In\'es de la Cruz 3, 28049 Cantoblanco, Madrid, Spain.}

\author{Guillermo Tajuelo-Castilla}
\affiliation{Instituto de Ciencia de Materiales de Madrid (ICMM. CSIC). Materials Science Factory. Structure of Nanoscopic Systems Group. c/ Sor Juana In\'es de la Cruz 3, 28049 Cantoblanco, Madrid, Spain.}

\author{Luis V\'azquez}
\affiliation{Instituto de Ciencia de Materiales de Madrid (ICMM. CSIC). Materials Science Factory. Structure of Nanoscopic Systems Group. c/ Sor Juana In\'es de la Cruz 3, 28049 Cantoblanco, Madrid, Spain.}

\author{Jes\'us M. Sobrado}
\affiliation{Centro de Astrobiolog\'i{}a (CAB. INTA-CSIC). Crta- de Torrej\'on a Ajalvir km4, 28850, Torrej\'on de Ardoz, Madrid, Spain.}

\author{Marcelino Ag\'undez}
\affiliation{Instituto de F\'\i{}sica Fundamental (IFF. CSIC). Group of Molecular Astrophysics. c/ Serrano 123, 28006 Madrid, Spain.}

\author{Miguel Jim\'enez-Redondo}
\affiliation{Instituto de Estructura de la Materia (IEM.CSIC). Molecular Physics Department. c/Serrano 123, 28006 Madrid, Spain.}

\author{V\'i{}ctor J. Herrero}
\affiliation{Instituto de Estructura de la Materia (IEM.CSIC). Molecular Physics Department. c/Serrano 123, 28006 Madrid, Spain.}

\author{Isabel Tanarro}
\affiliation{Instituto de Estructura de la Materia (IEM.CSIC). Molecular Physics Department. c/Serrano 123, 28006 Madrid, Spain.}

\author{Jos\'e Cernicharo}
\affiliation{Instituto de F\'\i{}sica Fundamental (IFF. CSIC). Group of Molecular Astrophysics. c/ Serrano 123, 28006 Madrid, Spain.}

\author{Jos\'e \'Angel Mart\'i{}n-Gago}
\affiliation{Instituto de Ciencia de Materiales de Madrid (ICMM. CSIC). Materials Science Factory. Structure of Nanoscopic Systems Group. c/ Sor Juana In\'es de la Cruz 3, 28049 Cantoblanco, Madrid, Spain.}

\begin{abstract}
Silicon is present in interstellar dust grains, meteorites and asteroids, and to date thirteen silicon-bearing molecules have been detected in the gas-phase towards late-type stars or molecular clouds, including silane and silane derivatives. In this work, we have experimentally studied the interaction between atomic silicon and hydrogen under physical conditions mimicking those at the atmosphere of evolved stars. We have found that the chemistry of Si, H and H$_2$ efficiently produces silane (SiH$_4$), disilane (Si$_2$H$_6$) and amorphous hydrogenated silicon (a-Si:H) grains. Silane has been definitely detected towards the carbon-rich star IRC\,+10216, while disilane has not been detected in space yet. Thus, based on our results, we propose that gas-phase reactions of atomic Si with H and H$_2$ are a plausible source of silane in C-rich AGBs, although its contribution to the total SiH$_4$ abundance may be low in comparison with the suggested formation route by catalytic reactions on the surface of dust grains. In addition, the produced a-Si:H dust analogs decompose into SiH$_4$ and Si$_2$H$_6$ at temperatures above 500 K, suggesting an additional mechanism of formation of these species in envelopes around evolved stars. We have also found that the exposure of these dust analogs to water vapor leads to the incorporation of oxygen into Si-O-Si and Si-OH groups at the expense of SiH moieties, which implies that, if this type of grains are present in the interstellar medium, they will be probably processed into silicates through the interaction with water ices covering the surface of dust grains. 

\end{abstract}


\section{Introduction}
Hydrogen and helium constitute more than 99$\%$ of the matter in the Universe. Among the trace elements, silicon is the fifth most abundant, being a usual component of the solid refractory material spread in interstellar and circumstellar clouds. Silicon has been identified in the dust grains formed around evolved stars through infrared emission bands, in the form of solid SiC in carbon-rich sources \citep{Treffers1974} and silicates, both amorphous and crystalline, around oxygen-rich stars \citep{Waters1996, Henning2010}. Silicates are also a major constituent of interstellar dust grains \citep{Fogerty2016}. To date, thirteen silicon-bearing molecules have been detected in gas-phase towards the carbon-rich star IRC\,+10216 and in a few other sources. SiO and SiS were the first species detected towards evolved stars and molecular clouds by comparison with their laboratory spectra \citep{Toerring1968,Tiemann1972}. Among the other silicon-bearing species, SiC \citep{Cernicharo1989}, SiC$_2$ \citep{Thaddeus84}, and Si$_2$C \citep{cernicharo2015} are the most promising candidates as seeds of SiC-dust in carbon-rich stars \citep{Massalkhi2018}. Other silicon-bearing species found in the envelopes of carbon-rich evolved stars are SiN \citep{Turner1992}, SiH$_3$CN \citep{Agundez2014,Cernicharo2017}, SiH$_3$CH$_3$ \citep{Cernicharo2017}, SiC$_3$ \citep{Apponi1999}, SiC$_4$ \citep{Ohishi1989}, SiCN \citep{Guelin2000}, and SiNC \citep{Guelin2004}. Silane (SiH$_4$) has been observed through the ro-vibrational transitions near 917 cm$^{-1}$ \citep{Goldhaber1984,Keady1993} towards the carbon-rich star IRC\,+10216. 

To investigate the formation of silicon-containing species in astrophysical environments, most of the laboratory experiments employed SiH$_4$ as molecular precursor. For instance, mixtures of silane and CO or C$_2$H$_2$ subjected to low pressure discharges efficiently produce SiO, SiC, SiC$_2$ and other SiC$_n$ molecules, already detected in space \citep{McCarthy2003}. Moreover, as discussed by \cite{Kaiser2005}, several authors studied the photochemistry of SiH$_4$ and the subsequent production of hydrogenated silicon clusters, such as SiH$_2$  and SiH, thus suggesting a rich silane chemistry occurring in the circumstellar envelopes. In fact, these radicals can react with the surrounding species (for instance CH$_4$) to form organosilicon molecules. Recently, also the interaction of the silyldyne (SiH) radical and excited atomic Si with hydrocarbons have been shown to lead to the formation of complex organosilicon molecules such as those ubiquitously found in the CSE of IRC+10216 \citep{Yang2015,Yang2019}. 

On the other hand, since the 1970s, the solar cell industry encouraged the study of the thermal decomposition and nucleation of silane, that is one of the methods to produce solar-grade silicon with application to solar cells. For instance \cite{Wyller2016} studied the particle nucleation from the decomposition of SiH$_4$ in a H$_2$ atmosphere as a function of temperature in a free space reactor. The formation of SiH$_4$ itself from its simplest components, i.e., Si and H$_2$, has been experimentally investigated by \cite{Hanfland2011} by compressing (124 GPa) elemental silicon and hydrogen at room temperature into a diamond anvil cell, i.e., at conditions far removed from those at the CSEs of AGBs.

In this work, we investigate the chemistry of atomic silicon and hydrogen, under conditions mimicking those reported for the circumstellar regions of asymptotic giant branch (AGB) stars, using the Stardust machine, an experimental station allowing for the synthesis, processing and in-situ characterization of the gaseous products and dust analogs produced from atomic precursors \citep{Martinez2020,Santoro2020}. We show that the interaction between Si atoms, H and H$_2$ efficiently generates silane (SiH$_4$), disilane (Si$_2$H$_6$), and amorphous hydrogenated silicon (a-Si:H) grains. Furthermore, we show that the thermal decomposition of the a-Si:H dust analogs produces SiH$_4$ and Si$_2$H$_6$ as well. The a-Si:H analogs also exhibit a high reactivity when exposed to water vapor, incorporating oxygen into Si-O and Si-OH moieties. We discuss the implications of our experimental results for the chemistry occurring in circumstellar envelopes around AGB stars.

\section{Experiments}

The synthesis and characterization of the products of Si + H$_2$ reaction have been performed with the Stardust machine, which has been thoroughly described elsewhere \citep{Martinez2018,Martinez2020,Santoro2020}. Briefly, this experimental station is based on the use of a sputtering gas aggregation source (SGAS) to extract atoms from a solid target by means of magnetron sputtering. Inside the aggregation zone of the SGAS, the atoms condense into molecules and nanoparticles. In addition, it is possible to inject gases inside the aggregation zone to promote gas-phase chemical reactions during nanoparticle formation. A battery of in-situ diagnostic techniques is available for characterizing both the gas-phase and solid-phase products. Moreover, a fast entry port allows the collection of solid samples for ex-situ analysis. The complete system is kept in ultra-high vacuum (UHV) with a base pressure of 2 $\times$ 10$^{-10}$ mbar, ensuring an ultra-clean environment for both the synthesis and characterization of the analogs.

In the present study, a polycrystalline silicon target (99.99$\%$ purity) was used to vaporize silicon atoms by radio-frequency magnetron sputtering, using argon as the sputtering gas with a flow rate of 150 sccm. The power applied to the magnetron was 100 W. A controlled flow of H$_2$ (99.99$\%$ purity) was injected into the aggregation zone at flow rates of 0, 0.15, 1 and 5 sccm. The solid dust analogs analyzed were produced using a flow rate of 1 sccm.

Optical emission spectroscopy (OES) of the sputtering plasma was performed inside the aggregation zone at a distance of around 1 cm from the magnetron target through a fused silica window. A Czerny-Turner spectrograph (Shamrock SR-193-i-A, Andor) with a CCD camera (iDus DU420A-BVF) was employed with a spectral resolution of 0.15 nm.

Mass spectrometry was carried out with a Pfeiffer HiQuad QMG 700 with QMA 400 mass spectrometer (mass range of 0 to 512 amu) and a CP 400 ion counter preamplifier.

Morphological characterization of the dust analogs was performed ex-situ by atomic force microscopy (AFM). The dust analogs were collected on SiOx substrates, and a Nanoscope IIIA (Veeco) AFM System was employed. All images were analyzed using the WSxM software \citep{Horcas2007}.

Fourier transform infrared (FTIR) spectroscopy was performed in-situ in transmission geometry. The dust analogs were deposited on KBr substrates and transferred to a separated UHV chamber using an UHV suitcase (P $<$ 5 $\times$ 10$^{-9}$ mbar). A Vertex 70 V (Bruker) instrument equipped with a liquid nitrogen cooled Mercury Cadmium Telluride (MCT) detector was employed. The spectral resolution was 2 cm$^{-1}$ and 256 scans were coadded for each spectrum. 

Thermal programmed desorption (TPD) experiments were carried out by heating the solid samples up to 650 K with a rate of 10 K/min and using the above-mentioned mass spectrometer to detect the desorbing species. For the TPD measurements the dust analogs were collected on a clean Ag (111) substrate.

Finally, X-ray photoelectron spectroscopy (XPS) of the dust analogs was performed in-situ using a Phoibos 100 1D electron/ion analyzer with a one-dimensional delay line detector. Freshly cleaved highly ordered pyrolytic graphite (HOPG) was used as substrate.

\section{Results}

\begin{figure}
        \centering
        \includegraphics[width=1\linewidth]{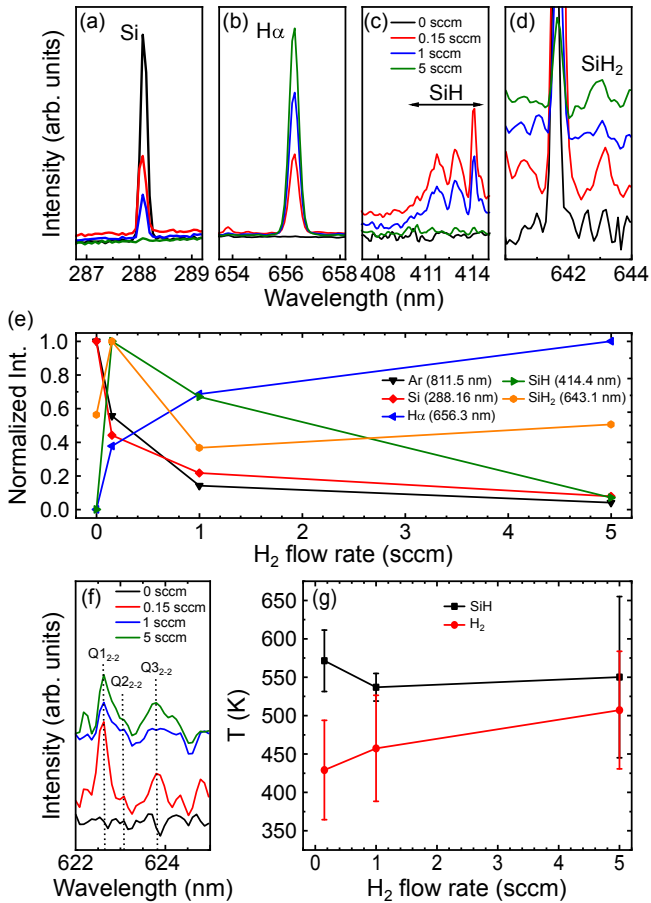}
        \caption{Optical emission spectra obtained during the experiment. The optical emission of Si (a), H$\alpha$ (b), SiH (c) and SiH$_2$ (d) are displayed for increasing H$_2$ flow rate. In (d), the curves have been vertically shifted for clarity. (e) Evolution of the Ar, Si, H$\alpha$, SiH and SiH$_2$ line intensities with the H$_2$ flow rate. The wavelength employed for each line is indicated in the figure. (f) Optical emission of the Q1$_{2-2}$, Q2$_{2-2}$ and Q3$_{2-2}$  ro-vibrational transitions of H$_2$ used to derive the gas temperature in the plasma. The curves have been vertically shifted for clarity. (g) Derived gas temperatures of SiH and H$_2$ at a distance of 1 cm from the magnetron target surface at the different H$_2$ flow rates used.}
        \label{fig:accolla_f1}
\end{figure}

When no H$_2$ is injected into the system, the presence of sputtered silicon atoms in the magnetron plasma is proven by OES through the line at 288.16 nm \citep{Striganov1968}. Figure~\ref{fig:accolla_f1}a shows that the signal originated from excited Si atoms is more intense in the absence of hydrogen. Injecting H$_2$ promotes an efficient chemistry, and the emission from excited Si atoms decreases as the amount of H$_2$ increases, nearly vanishing at a flow rate of 5 sccm (Figure~\ref{fig:accolla_f1}a), whereas the emission of atomic hydrogen (H$\alpha$) scales up with the H$_2$ flow rate (Figure~\ref{fig:accolla_f1}b), indicating that Si atoms are consumed in the reaction with hydrogen. Figure~\ref{fig:accolla_f1}c shows the detection of SiH \citep{Perrin1980} at a flow rate of 0.15 sccm, whose emission decreases as we raise the amount of hydrogen, vanishing when the H$_2$ flow rate reaches 5 sccm.
In addition, a faint emission line is observed at 643.10 nm which can be attributed to the presence of SiH$_2$ \citep{Escribano1998}, whose maximum intensity corresponds as well to a H$_2$ flow rate of 0.15 sccm (Fig.~\ref{fig:accolla_f1}d). An Ar emission line at 643.15 nm overlaps with the emission of SiH$_2$. However the different behavior of Ar emission and the line at 643.10 nm \textit{vs.} the H$_2$ flow rate (Fig.~\ref{fig:accolla_f1}e) supports our assignment. The decrease in Ar intensity with the introduction of reactive gases in the aggregation zone of ion cluster sources is a known process and can be ascribed to the consumption of part of the power applied to the magnetron in the dissociation the reactive gas molecules \citep{Martinez2018}. The behavior of SiH and SiH$_2$ with respect to the H$_2$ flow rate, suggests a critical concentration of H$_2$ to trigger an active chemistry.

We estimate that around 10$\%$ of the injected H$_2$ reaches the magnetron, being therefore susceptible to dissociate in the plasma, which produces H atoms that are available to react with Si atoms. The bimolecular reaction of Si in its electronic ground state with H$_2$ to form SiH + H is endothermic by about 35 kcal/mol \citep{Zanchet2018} and therefore does not take place at the temperatures involved in our experiment (see below). However, to date, no data are available for the radiative association reaction Si + H $\to$ SiH, nor for the bimolecular reaction of electronically excited Si with H$_2$ to form SiH + H, though Si($^1$D) has been shown to be highly reactive with, e.g., hydrocarbons \citep{Yang2019}. Our results from OES suggest that excited Si atoms are consumed in the reaction with H to form SiH and, therefore, in our case, since radiative association reactions are very slow for these species, the chemistry is initiated by the three-body reaction Si + H + Ar $\to$ SiH + Ar, in which Ar (the sputtering gas) accelerates the chemistry. Moreover, the bimolecular reaction of electronically excited Si with H$_2$ to form SiH + H might contribute as well. Once SiH is formed, SiH$_2$ is produced through SiH + H + Ar$\to$ SiH$_2$ + Ar, as suggested by the behaviour of SiH and SiH$_2$ with respect to the H$_2$ flow rate (Fig.~\ref{fig:accolla_f1}e).

From the OES measurements, we have also derived the rotational temperatures of H$_2$ and SiH, which, in the case of equilibrium with the translational temperature, can be considered as the gas temperature in the plasma \citep{Shimada2006}. This condition is met in magnetron sputter sources since the pressure of the sputtering gas employed guarantees enough collisions for the rotational and translational temperatures to be in equilibrium \citep{RenHow2018}. In the case of SiH we have simulated the emission spectra using the LIFBASE software \citep{Luque1999} at several rotational temperatures whereas in the case of H$_2$, we have obtained the rotational temperature applying the Boltzmann plot method to the Q-branch ($\Delta$J=0) ro-vibrational transitions Q1$_{2-2}$ ($\lambda$=622.58 nm), Q2$_{2-2}$ ($\lambda$=623.06 nm) and Q3$_{2-2}$ ($\lambda$=623.82 nm) observed within the Fulcher-$\alpha$ (d$^{3}\Pi_{u}$ – a$^{3}\Sigma^{+}_g$) band \citep{Shikama2007} (Fig.~\ref{fig:accolla_f1}f). Rotational temperatures of around 550 K for SiH and 470 K for H$_2$ have been derived at all the H$_2$ flow rates employed (Fig.~\ref{fig:accolla_f1}g). These temperatures have been obtained from measurements at a distance of 1 cm from the target surface and, therefore, the gas temperature has already decreased by collisions in the plasma expansion, implying that the gas temperature closer to the sputter target surface might be higher.

\begin{figure}
        \centering
        \includegraphics[width=1\linewidth]{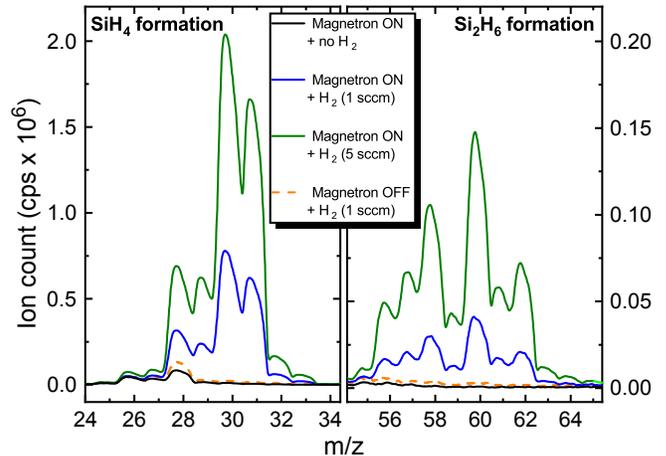}
        \caption{Mass spectra of the gas-phase products formed during the experiments. The increase of the peaks between m/z=28 and m/z=33 demonstrates silane (SiH$_4$) formation, whereas the increase of the peaks between m/z=56 and m/z=62 demonstrates disilane (Si$_2$H$_6$) formation.}
        \label{fig:accolla_f2}
\end{figure}

Figure~\ref{fig:accolla_f2} shows the mass spectrometry results for gas phase. In the absence of H$_2$, only residual gas traces are detected, indicating that, within the sensitivity of the instrument, no gas-phase species are formed. Once H$_2$ is injected, we detect a number of peaks ranging from m/z=28 to m/z=33 and from m/z=56 to m/z=62, which correspond to the electron-impact dissociation patterns of silane \citep{Kramida2019} and disilane \citep{SIMON1992}, respectively. Moreover, Figure ~\ref{fig:accolla_f2} shows that the amount of both molecules scales up with the H$_2$ flow rate.
Hence, SiH and SiH$_2$ radicals, detected by OES once H$_2$ is injected into the chamber, are the primary products that further react with H and/or H$_2$ to give rise to SiH$_4$ and Si$_2$H$_6$. In our experimental conditions, the possible reactions leading to the formation of silane are:

\medskip
\centerline{SiH + H $\to$ SiH$_2$  \citep{Raghunath2013}}
\medskip
\centerline{SiH + H$_2$ $\to$ SiH$_3$ \citep{Walch2001}}
\medskip
\centerline{SiH$_2$ + H $\to$ SiH$_3$ \citep{Raghunath2013}}
\medskip
\centerline{SiH$_3$ + H $\to$ SiH$_4$ \citep{Takahashi1994}}
\medskip
\centerline{SiH$_2$ + H$_2$ $\to$ SiH$_4$ \citep{Adamczyk2010}}
\medskip 
and to the formation of disilane:

\medskip
\centerline{SiH$_3$ + SiH$_3$ $\to$ Si$_2$H$_6$ \citep{Matsumoto1996}}
\medskip
\centerline{SiH$_2$ + SiH$_4$ $\to$ Si$_2$H$_6$ \citep{Matsumoto2005}}
\medskip

\begin{figure}
        \centering
        \includegraphics[width=1\linewidth]{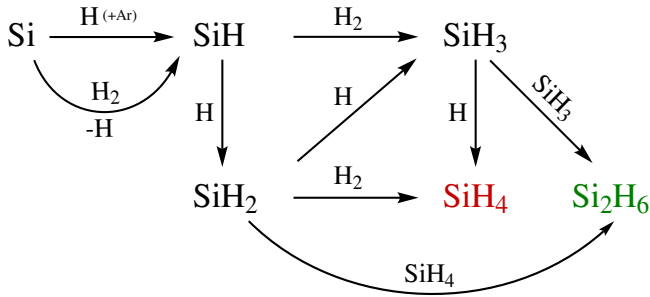}
        \caption{Scheme of the chemical reactions involved in the formation of SiH$_4$ and Si$_2$H$_6$. In our case the chemistry is initiated by the three-body reaction Si + H + Ar $\to$ SiH + Ar. Moreover, the bimolecular reaction Si + H$_2$ $\to$ SiH + H with atomic Si in an electronically excited state might contribute as well. Subsequently, since radiative association reactions for these species are slow, in our experiments the growth of Si$_n$H$_m$ species proceeds through three-body reactions with Ar (the sputtering gas).}
        \label{fig:accolla_s1}
\end{figure}

which are summarized in Figure~\ref{fig:accolla_s1}. All the radiative association reactions listed above are very slow and, therefore, in our case Ar acts as third body accelerating the chemistry.

\begin{figure}
        \centering
        \includegraphics[width=1\linewidth]{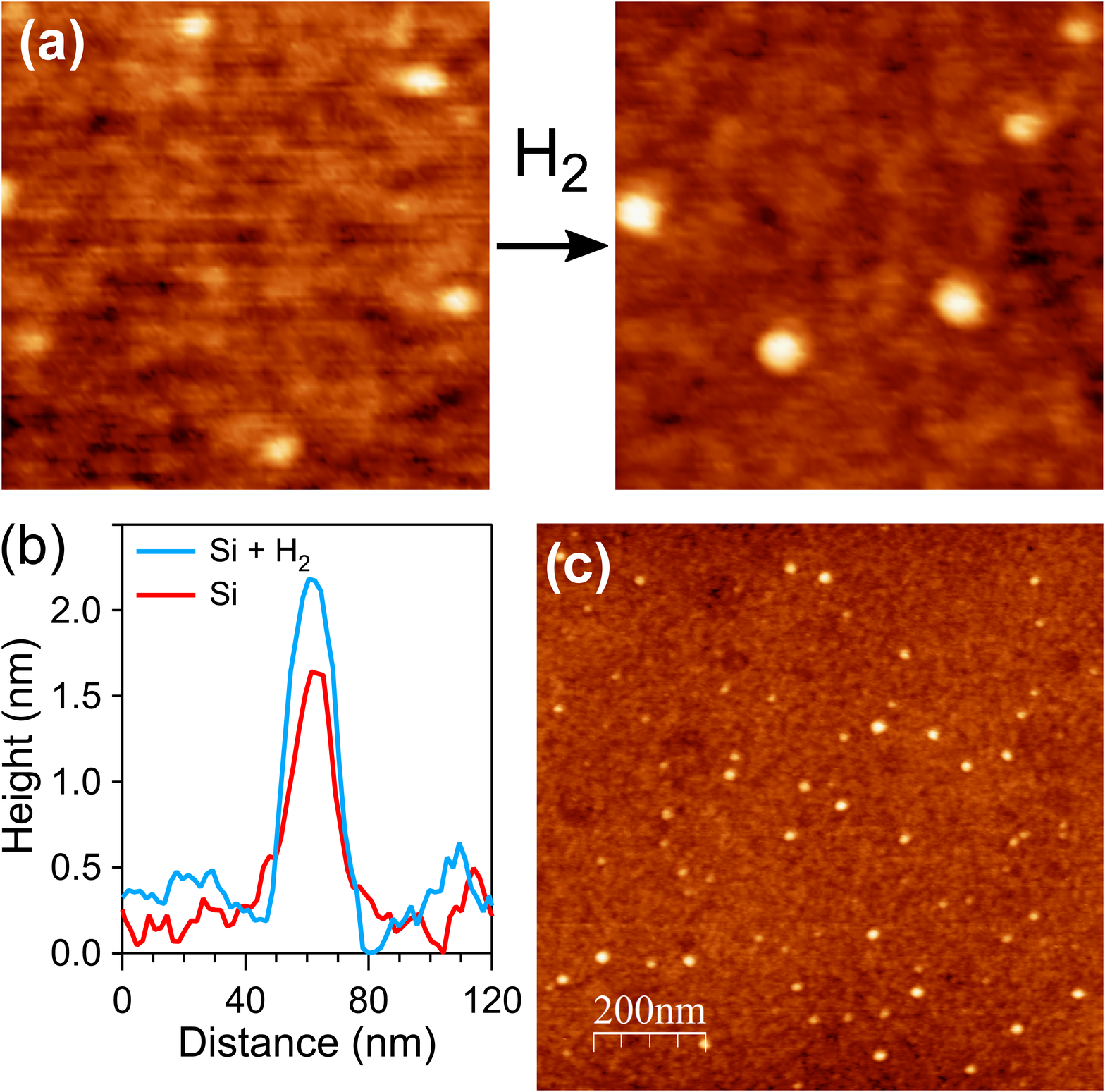}
        \caption{(a) AFM images (200 $\times$ 200 nm$^2$) of silicon nanoparticles: pure Si (left) and Si + H$_2$ (1 sccm) (right). (b) The height profiles of the nanoparticles show the increase in size after H$_2$ injection. (c) Lower magnification image of nanoparticles prepared following the reaction Si + H$_2$ (1 sccm).}
        \label{fig:accolla_f3}
\end{figure}

Apart from the gas phase products, the interaction of silicon and hydrogen also produced solid dust analogs that consist of spherical nanoparticles, as shown in the AFM images of Figure~\ref{fig:accolla_f3}. In the absence of H$_2$, Si atoms ejected from the magnetron aggregate forming solid nanoparticles (Figure~\ref{fig:accolla_f3}a), although their formation rate is very low. Injection of hydrogen into the system boosts the particle size and production efficiency (Figure~\ref{fig:accolla_f3}c). The particle formation rate increases more than 30 times, as derived from substrate coverage measurements, and the average size of the grains increases upon hydrogen introduction into the system from 1.5 $\pm$ 0.2 nm (pure Si chemistry) to 2.1 $\pm$ 0.2 nm (H$_2$ injected in the chamber), extracted from statistical analysis of the AFM images. Figure~\ref{fig:accolla_f3}b shows typical height profiles of the nanoparticles produced with and without hydrogen.

\begin{figure}
        \centering
        \includegraphics[width=1\linewidth]{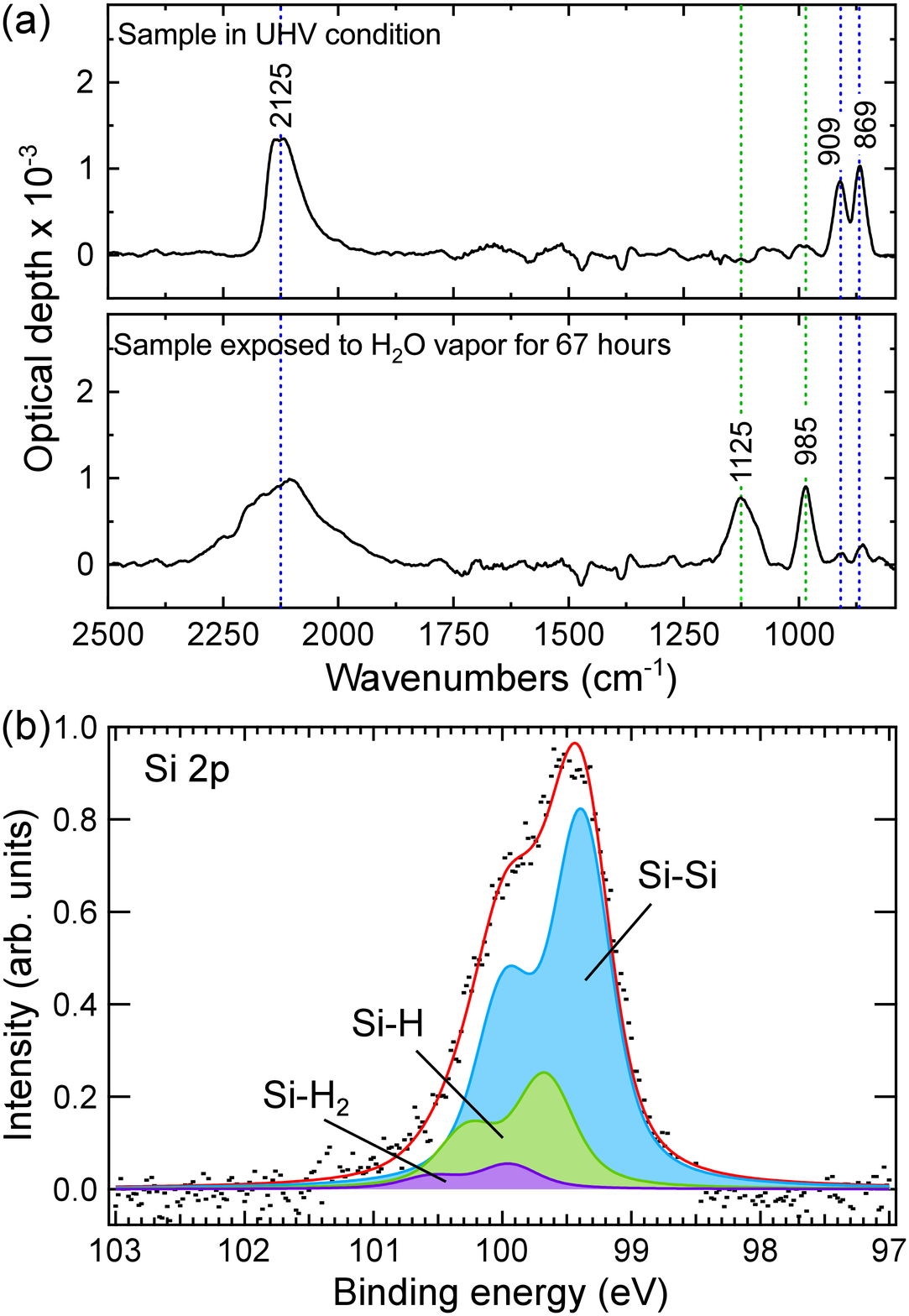}
        \caption{(a) FTIR spectra of the solid material synthesized following the Si + H$_2$ (1 sccm) reaction. Top panel: spectrum of the as-grown sample. Bottom panel: spectrum of the same sample after water vapor exposure for 67 hours. (b) High-resolution XPS scan for the Si 2p core level peak of the solid material synthesized following the Si + H$_2$ (1 sccm) reaction along with its deconvolution into components. Black dots: experimental data points. Red line: best fit.}
        \label{fig:accolla_f4}
\end{figure}

The solid dust analogs have also been characterized using three complementary characterization techniques: FTIR, XPS, and TPD. The infrared spectrum of the dust analogs shown in Figure~\ref{fig:accolla_f4}a exhibits a band centered at 2125 cm$^{-1}$ and a doublet (909 and 869 cm$^{-1}$) attributable, respectively, to SiH$_x$ stretching and SiH$_x$ bending/scissoring modes \citep{Stryahilev2000}. A FTIR spectrum of a fresh sample exposed to a controlled amount of water vapor (partial pressure equal to 1.0 $\times$ 10$^{-6}$ mbar during 67 hours) is shown in the lower panel of Figure~\ref{fig:accolla_f4}a. The intensity of the SiH$_x$ bending modes strongly decreases and two new features emerge at 1125 and 985 cm$^{-1}$, attributable respectively to Si-O-Si stretching \citep{Sabri2014} and to Si-OH bending \citep{Carteret2006} vibrations. Moreover, the shape and the intensity of the 2125 cm$^{-1}$ band is modified after water vapor exposure; in particular, the broadening of this band toward higher wavenumbers is indicative of silicon oxidation \citep{Moore1991}.

In Figure~\ref{fig:accolla_f4}b, we present a high-resolution XPS spectrum of the Si 2p peak. Comparing this peak with the well-determined binding energy for Si 2p in Si-Si configuration (99.5 eV), we can estimate the hydrogen content of the nanoparticles. Silicon oxide components, expected at 102-104 eV, are not present, indicating the absence of oxidation. The peak can be deconvolved into three doublet components, each doublet arising from spin-orbit splitting with $\Delta$=0.6 eV \citep{Lu1993}. Each component is a signature of Si atoms in the nanograins with a distinct chemical/electronic environment. The Si 2p$_{3/2}$ core-level has three components whose maxima correspond to binding energies of 99.4, 99.7 and 100.0 eV (blue, green and purple components in Figure~\ref{fig:accolla_f4}b respectively), which can be readily assigned to Si-Si, Si-H and Si-H$_2$ species respectively \citep{Cerofolini2003}. From the relative intensities of the components, we can estimate the relative abundance of Si atoms in every bonding configuration. The intensity analysis yields 73$\%$ in Si-Si, 23$\%$ in Si-H, and 5$\%$ in Si-H$_2$.

\begin{figure}
        \centering
        \includegraphics[width=1\linewidth]{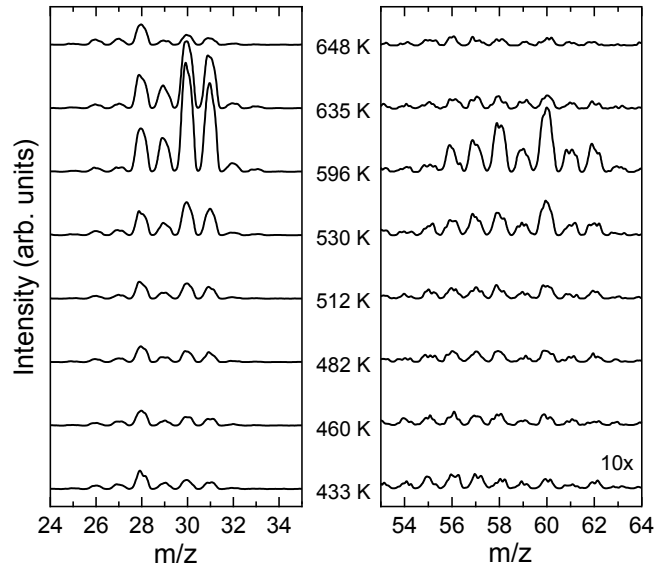}
        \caption{Thermal desorption of silane and disilane detected by the mass spectrometer during the annealing of the solid material synthesized following the Si + H$_2$ (1 sccm) reaction. The curves have been vertically shifted and the signal from m/z=53 to m/z=64 has been magnified 10 times for the sake of clarity.}
        \label{fig:accolla_f5}
\end{figure}

Finally, we have characterized the dust grains by means of TPD. Figure~\ref{fig:accolla_f5} shows clear desorption of silane (m/z = 28 - 33) and disilane (m/z = 56 - 62) from the solid-state material within the temperature range 530 - 648 K, being the intensity of disilane about 15 times lower than that of silane.

All the experimental results suggest that the dust analogs are composed of amorphous hydrogenated silicon (a-Si:H) \citep{Street1991}. This material consists of a random network of Si atoms in sp$^3$ hybridization and covalently bonded to four immediate neighbors, which can be either Si or H atoms exclusively. Amorphous hydrogenated silicon can be grown in thin films by magnetron sputtering of Si targets in the presence of hydrogen \citep{Langford1992} with a final hydrogen content ranging between 8 and 45 $\%$ \citep{Street1991}. In addition, annealing of a-Si:H is known to produce major structural changes in the material, with massive outgassing starting above 625 K, corresponding to conversion of SiH$_3$ groups to SiH$_2$ and SiH \citep{Langford1992}. In our case, a-Si:H is grown in grains with a hydrogen content of about 15 $\%$ as derived from XPS measurements. As the material is heated during the TPD experiments, the highly reactive silicon-hydrogen species produced during the thermal processing of the grains react with trapped H$_2$ and between them to form the more stable saturated molecules silane and disilane. Note that the IR spectrum of Figure~\ref{fig:accolla_f4}a contains the same bands as those of a-Si:H \citep{Langford1992}.

\section{Discussion}

Our experiments evidence that atomic Si and hydrogen efficiently react in gas-phase to form silane, disilane and amorphous hydrogenated silicon nanoparticles. As monitored by OES, the synthesis of the two gaseous species is mediated by the formation of SiH and SiH$_2$. The former is produced by the three-body reaction Si + H + Ar $\to$ SiH + Ar, which involves the sputtering gas, and the bimolecular reaction of electronically excited Si with H$_2$ to form SiH + H might contribute as well. Once SiH is produced, in our case SiH$_2$ forms from the reaction of SiH  and H with Ar as third body, since radiative association reactions are slow for these species. Subsequently, SiH and SiH$_2$ can react with atomic and molecular hydrogen to form the saturated Si compounds (Si$_n$H$_{2n+2}$) through three-body reactions in which Ar acts as third body accelerating the chemistry. 

From the astrochemical perspective, silicon and hydrogen have high cosmic abundances and thus silane and silane derivatives are candidates to be present in interstellar and circumstellar environments. To date, silane has been firmly detected only in the circumstellar envelope of the carbon-rich AGB star IRC\,+10216 \citep{Goldhaber1984}, whereas disilane has never been detected in space. The importance of silane chemistry in the CSE of IRC+10216 is stressed by the detection of methyl silane and silyl cyanide by \cite{Cernicharo2017}. However, the radius at which silane is formed in IRC+10216 is not well constrained. \cite{Keady1993} suggest that it could be formed at 40 stellar radii based on the analysis of the profiles of several ro-vibrational lines, which set an upper temperature limit of about 280 K for silane formation, whereas \cite{Monnier2000} claim that the formation of silane occurs at $\sim$ 80 stellar radii, based on interferometric observations of one of its ro-vibrational lines. The derived abundance of silane (around 2 $\times$ 10$^{-7}$ relative to H$_2$) is larger than predicted by thermochemical equilibrium by several orders of magnitude \citep{Gail2013,Agundez2020} and thus it has been suggested that SiH$_4$ could be formed by catalytic reactions at the surface of dust grains \citep{Keady1993,Monnier2000}. On the other hand, chemical calculations including the effect of shock-waves in the inner region of IRC+10216 predict that SiH$_4$ can be formed at 5 stellar radii, although at low abundances \citep{Willacy1998}.

Our results present a mechanism to form silane from atomic silicon and molecular hydrogen in the innermost regions of the CSE of C-rich AGBs like IRC\,+10216. Thermochemical equilibrium predicts that atomic silicon is very abundant in the inner regions (up to $\sim$ 5 stellar radii) of carbon-rich envelopes, although at larger distances SiS, SiO, and silicon carbides are predicted to trap most of the silicon \citep{Gail2013,Agundez2020}. Nevertheless, there are evidences of non thermal equilibrium in the inner wind of AGBs, for instance the ascertained presence of a few molecules not expected to form under thermal equilibrium \citep{Agundez2020}. Thus, models based on non-equilibrium chemistry predict that atomic silicon can be injected into carbon-rich expanding envelopes with fairly large abundances \citep{Cherchneff2006}, and its abundance, for carbon stars, increases with radius. Moreover, other refractory elements, such as Na, K, Ca, Cr, and Fe, have been observed to be present in atomic form in the outer envelope of IRC\,+10216 \citep{Mauron2010}. However, our experiments indicate that three-body reactions are needed to form SiH$_x$ (x=1-4) species from atomic Si and atomic and molecular hydrogen at sufficient reaction rates and, thus, the gas-phase synthesis of SiH$_4$ is only possible at distances lower than 10 stellar radii, even if atomic Si is present in the outer envelope. Our results also point towards the gas-phase synthesis of disilane in the inner envelope and the work of \cite{Kaiser2005} could guide the search for Si$_2$H$_x$ species through infrared observations.

As above mentioned, the formation of SiH$_4$ in IRC+10216 is thought to start at 40 stellar radii, suggesting that the main contribution to the formation of silane might come from chemical reactions catalyzed on the surface of dust grains\citep{Keady1993} with the condensation of  Si-bearing molecules on the grains and surface diffusion of those playing major a role \citep{Monnier2000}. Likely, these are the main mechanisms on SiH$_4$ formation in IRC+10216 since, if gas-phase synthesis would be the main mechanism, the abundance of SiH$_4$ would follow a different spatial distribution. Nevertheless, based on our results, a contribution from the gas phase synthesis, even if small, might start earlier in the expanding wind of IRC+10216. The fact that SiH$_4$ is not detected at distances lower than 40 R* indicates that the gas-phase reactions might be slow with lower reaction rates than other competing reactions of atomic Si with, e.g., C$_2$H$_2$.

In addition to the products formed in the gas-phase, in our experiments we have generated amorphous hydrogenated silicon grains, which consist of a random three-dimensional network in which silicon atoms are tetragonally bonded to other silicon and hydrogen atoms. In fact, both XPS and FTIR spectroscopies indicate that hydrogen atoms are bound to the silicon atoms. TPD experiments show that these nanograins decompose at temperatures above 530 K leading to silane and disilane. To date, there are not observational evidences of the presence of solid hydrogenated silicon in space. Whether this kind of grains are present in interstellar and circumstellar medium is still unclear and the fact that this material is more thermally unstable than other Si-containing condensates plays against its presence in space since the formation of dust in envelopes around AGB stars takes place as the expanding gas cools down to temperatures below 1500-1000 K. In this temperature range, Si-containing solids like SiC (in the case of C-rich envelopes) and silicates (in O-rich environments) condense. Therefore, it is likely that at lower temperatures, where a-Si:H can form without thermally decomposing, most silicon has been already locked by more refractory condensates. However, we note that since dust formation in AGB ejecta occurs out of equilibrium \citep{Gail2013}, the way in which silicon condenses does not need to strictly follow the condensation sequence given by thermochemical equilibrium. For instance, condensates like MgS, which have fairly low condensation temperatures ($<$ 600 K; \citealt{Lodders1999,Agundez2020}), have been proposed to be part of the dust grain composition in carbon-rich AGB stars \citep{Goebel1985}. Thus, if hydrogenated silicon grains form in carbon-rich circumstellar envelopes, processing by ultraviolet photons and/or cosmic rays penetrating into intermediate layers of the envelope might lead to the formation of silane. Likely, this is not the main mechanism for silane formation, but it could contribute to justify its amount, 6 times higher than the one predicted by models based on thermodinamic equilibrium.

Finally, our experiments have also shown the conversion of Si-H bonds to Si-O-Si and Si-OH bonds upon exposure of the hydrogenated silicon nanoparticles to water vapor. Si-H bonds are known to exhibit a high reactivity with oxygen bearing species \citep{Nuth1992}. If hydrogenated silicon is present in the interstellar medium, in the colder regions where dust grains are coated by icy species, mainly H$_2$O, it is likely that it be oxidized forming silicates.

\section{Conclusions}

The experimental results presented here provide a possible and efficient mechanism for the formation of silane and disilane in the gas phase from Si, H and H$_2$ in the innermost regions of the CSE around AGB stars. Most likely, the main formation mechanism of SiH$_4$ in C-rich stars, as suggested by observations, involves the catalysis on the surface of dust grains since gas-phase reactions are only efficient in the inner parts of the envelope and cannot explain the observed abundances of silane and its derivatives in the CSE of IRC+10216. In our experiments, hydrogenated silicon dust (a-Si:H) particles are also formed, the thermal processing of which produces SiH$_4$ and Si$_2$H$_6$ at temperatures above 530 K. This kind of grains could form and survive without thermally decomposing in regions of AGB envelopes where temperatures are below $\sim$500 K and, if exposed to energetic photons or particles, could become a source of silane. In addition, we have observed that the exposure of the a-Si:H dust particles to water vapor promotes the formation of SiO and SiOH groups at the expense of SiH moieties. Therefore, if hydrogenated silicon dust is coated with water ices in the interstellar medium, it will most likely turn into silicate grains.

\acknowledgments

We thank the European Research Council for funding support under Synergy grant ERC-2013-SyG, G.A. 610256 (NANOCOSMOS). Also, we acknowledge partial support from the Spanish MINECO through grants MAT2017-85089-c2-1R, FIS2016-77726-C3-1-P, FIS2016-77578-R, AYA2016-75066-C2-1-P and RyC-2014-16277. Support from the FotoArt-CM Project (P2018/NMT 4367) through the Program of R$\&$D activities between research groups in Technologies 2013, co-financed by European Structural Funds, is also acknowledged. G.TC. acknowledges funding from the Comunidad Aut\'noma de Madrid (PEJD-2018-PRE/IND-9029).

\bibliography{ApJLett_Accolla_final.bib}{}
\bibliographystyle{aasjournal}

\end{document}